\documentclass[reprint,
superscriptaddress,
 amsmath,amssymb,
 aps,
 prl,longbibliography,maxbibnames=10]{revtex4-2}

\usepackage{graphicx}
\usepackage{csquotes}
\usepackage{graphicx}
\usepackage{booktabs}
\usepackage{siunitx}
\usepackage{xcolor}
\usepackage{ulem}
\usepackage{pdfpages}
\makeatletter
\AtBeginDocument{\let\LS@rot\@undefined}
\makeatother

\usepackage{fp}
\newcount\boxheight
\newcount\boxwidth
\newcommand\testaspect[1]{%
 \setbox0=\hbox{#1}%
 \boxheight=\ht0\relax%
 \boxwidth=\wd0\relax%
 \FPdiv\theaspect{\the\boxheight}{\the\boxwidth}%
 \global\let\theaspect\theaspect%
 \copy0%
}

\begin{document}

\title{Interferometric imaging using shared quantum entanglement}

\author{Matthew R. Brown}
\affiliation{Department of Physics and Oregon Center for Optical, Molecular, and Quantum Science, University of Oregon, Eugene, Oregon 97403, USA}
\author{Markus Allgaier}
\affiliation{Department of Physics and Oregon Center for Optical, Molecular, and Quantum Science, University of Oregon, Eugene, Oregon 97403, USA}
\author{Val\'erian Thiel}
\affiliation{PASQAL, 7 rue L\'eonard de Vinci, 91300 Massy, France}
\author{John D. Monnier}
\affiliation{Department of Astronomy, University of Michigan, Ann Arbor, Michigan 48109, USA}
\author{Michael G. Raymer}
\affiliation{Department of Physics and Oregon Center for Optical, Molecular, and Quantum Science, University of Oregon, Eugene, Oregon 97403, USA}
\author{Brian J. Smith}
\email{bjsmith@uoregon.edu}
\affiliation{Department of Physics and Oregon Center for Optical, Molecular, and Quantum Science, University of Oregon, Eugene, Oregon 97403, USA}

\date{\today}

\begin{abstract}
Quantum entanglement-based imaging promises significantly increased resolution by extending the spatial separation of optical collection apertures used in very-long-baseline interferometry for astronomy and geodesy. We report a table-top entanglement-based interferometric imaging technique that utilizes two entangled field modes serving as a phase reference between two apertures. The spatial distribution of a simulated thermal light source is determined by interfering light collected at each aperture with one of the entangled fields and performing joint measurements. This experiment demonstrates the ability of entanglement to implement interferometric imaging.
\end{abstract}

\maketitle

Coherent measurement of light entering separate collection apertures of an imaging system, an approach called aperture synthesis that forms the basis of interferometric imaging, enables increased angular resolution beyond the single-aperture diffraction limit, as first demonstrated by Michelson and Pease  \cite{mich1921}. The resolution of interferometric imaging is limited, in-principle, only by the aperture separation (the ‘baseline’). Unlike interferometric imaging in the radio frequency (RF) band, where Earth-sized telescope arrays have been implemented based on locally recording the RF fields at each telescope \cite{radio}, in the optical band (visible and near-infrared) the maximum baseline length is limited, in part, by the difficulty in performing local coherent detection with low noise at the few-photon level. The desire to increase angular resolution in the optical band by increasing the baseline is motivated by a number of applications currently limited by resolution: the search for exoplanets and the study of their atmospheres, resolved imaging of black hole event horizons in the near-infrared to complement the mm-wave imaging of the Event Horizon Telescope (EHT), geodesy, imaging of planet-forming disks, and stellar surfaces beyond the Sun \cite{EHT2019,EHT2019II,EHT2022,schuh2012vlbi,PlanetFormation,Mon2007}. This has prompted the quantum information science community to search for new tools such as shared optical entanglement between the receivers to extend optical interferometric imaging baselines \cite{GJC,Tsang,khabiboulline2019I,khabiboulline2019II,howard2019,Kok,diaz2021,Czupryniak,Czupryniak2022}.   

\begin{figure}
    \centering
    \includegraphics[width=\columnwidth]{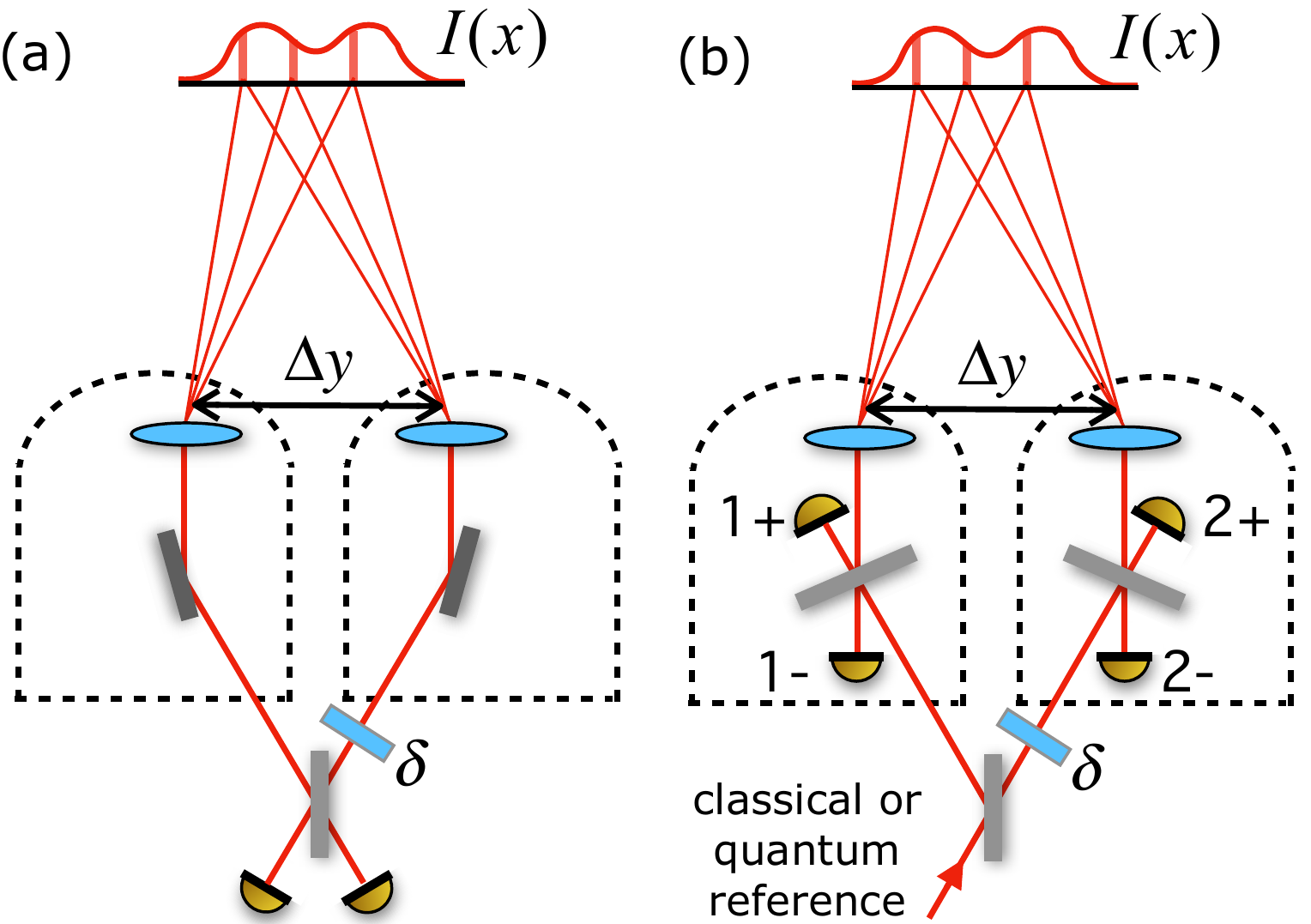}
    \caption{Schematics of two approaches to interferometric imaging. Light from a distant incoherent source with intensity distribution $I(x)$ is collected by apertures with baseline $\Delta y$. a) Direct interference combines the collected fields at a beam splitter with a known variable phase shift, $\delta$, and performs intensity measurements at the output. b) Indirect interference combines the collected fields with reference fields (laser light or path-entangled single-photon state) followed by local measurements at each aperture. Correlations between measurement outcomes at each telescope yield the mutual coherence function of the fields collected by the two telescope stations.}
    \label{fig1}
\end{figure}


There are three common approaches to achieving high angular resolution with synthetic apertures: 1) transporting the received fields to a central location and interfering them (direct interference) as depicted in Fig. \ref{fig1}(a); 2) distributing and interfering strong mutually coherent fields, known as local oscillators (LO), with two received fields, as shown in Fig. \ref{fig1}(b); or, 3) measuring intensity correlations between the collected fields as in the Hanbury Brown-Twiss (HBT) experiment \cite{HBT1,HBT2,HBT3,stankus2020}. In the first approach, the challenge to increasing the baseline lies in constructing low-loss optical channels between telescopes, with the current state-of-the-art baseline being \SI{330}{\m} long \cite{Chara330m}. The second approach is limited by the photon-number uncertainty of the LO (also called shot noise) as noted by Townes \cite{tow00}. In the optical regime, direct interference is typically preferred due to its higher signal-to-noise ratio (SNR) compared to interference with a classical LO \cite{hale2000,HetVsDirect}. In contrast, radio-frequency interferometric imaging (e.g., EHT) uses LOs since the effect of shot noise is significantly reduced: for the same average energy as one photon in the optical field, the RF field contains approximately $10^6$ photons \cite{tow00}. The third (HBT) approach is limited to relatively bright sources since it requires detection of at least two photons from the astronomical source, whereas the other methods require only one photon.

In this Letter, we present an experimental realization of an interferometric imaging technique that utilizes quantum entanglement as proposed by Gottesman, Jennewein, and Croke (GJC) \cite{GJC}. To increase the baseline while delivering high SNR at optical wavelengths, GJC proposed the use of a distributed, path-entangled single-photon state as a shared phase reference in conjunction with a quantum repeater network to alleviate transmission losses [3]. A path-entangled reference state (PERS) can be generated by, e.g., splitting a single photon occupying a single traveling mode into two paths of (un)equal length using a beam splitter (BS) \cite{oliver1989,tan1991,vanEnk2005}. The entangled state of the two modes (1 and 2) in the Fock basis is

\begin{equation}
    \label{eq}
    |\psi_{PERS}\rangle=\frac{1}{\sqrt{2}}\left( |1\rangle_1|0\rangle_2+e^{i\delta}|0\rangle_1|1\rangle_2\right).
\end{equation}

\noindent In the present application, the state provides a phase reference, $\delta$, between two locations while minimizing the shot noise associated with the reference field.

The two paths are interfered with light collected by the apertures, as depicted in Fig. \ref{fig1}(b). A coincidence detection event between photon-counting detectors at each aperture, labeled $\pm1$ and $\pm2$ in Fig. \ref{fig1}(b), realizes a quantum joint measurement allowing coherent detection. For weak thermal-like sources, schemes using joint quantum measurements of the received fields (e.g., direct interference or the GJC protocol) can yield more information per received photon than any scheme using independent measurements at each receiver (e.g., using classical LOs) [15]. Here independent measurements are defined generally to include local operations and classical communication.

\begin{figure*}
    \centering
    \includegraphics[width=\textwidth]{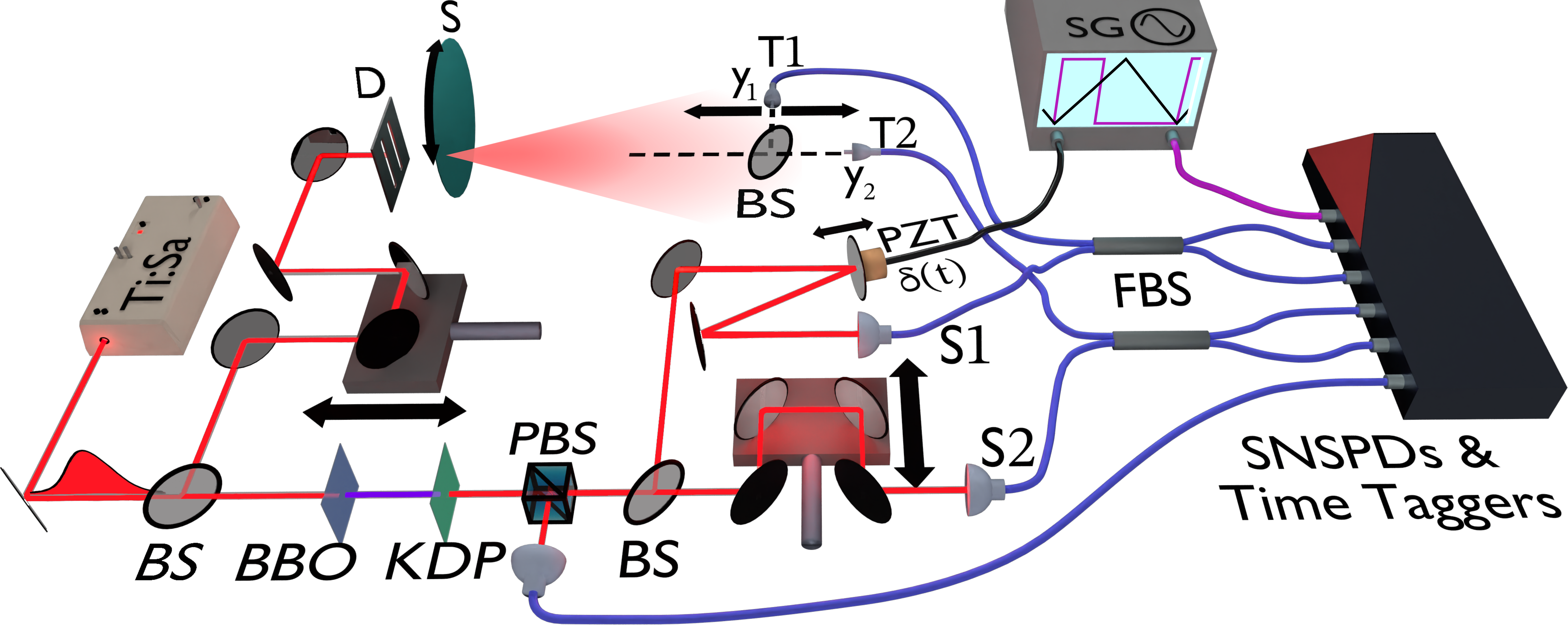}
    \caption{A titanium-sapphire (Ti:Sa) laser is separated into two paths by a beam splitter (BS). One path creates a single spectral-temporal-mode thermal-like state by passing through a double slit (D) and then a time-varying scatterer (S). The scattered light is collected by fibers T1 and T2 after a BS. The other path undergoes second-harmonic generation in a BBO crystal followed by parametric down conversion in a KDP crystal. The generated photon pair is split at a polarizing BS (PBS); one is used for heralding. The heralded photon is sent to a BS to generate the path-entangled reference state along two paths, with one path having a time-dependent length controlled by a PZT to introduce a phase shift, $\delta(t)$, driven by a signal generator (SG). Each path is coupled into single-mode fibers (S1 and S2), and combined with the fields collected by T1 and T2 in two fiber BS (FBS). The outputs of the FBSs are monitored by superconducting nanowire single-photon counting detectors (SNSPDs) and time taggers.}
    \label{fig2}
\end{figure*}


Interferometric imaging is based on measurement of the complex degree of coherence of an electromagnetic field after propagation from a source \cite{ManAndWolf}. In the detection plane, the complex degree of coherence is given by $j(\vec{y}_1,\vec{y}_2)=\langle E^*(\vec{y}_1)E(\vec{y}_2)\rangle/\sqrt{\langle |E(\vec{y}_1)|^2\rangle\langle |E(\vec{y}_2)|^2\rangle}$, where the electric field, $E$, is evaluated at two points labeled by transverse position vectors, $\vec{y}_1$ and $\vec{y}_2$, and angular brackets imply ensemble averaging. The complex visibility, $v(\Delta\vec{y})e^{i\phi(\Delta\vec{y})}$, arising from the interference of the fields collected at $\vec{y}_1$ and $\vec{y}_2$ is equal to $j(\vec{y}_1,\vec{y}_2)$. The van Cittert-Zernike theorem for a spatially incoherent source parallel to the detection plane shows that $v(\Delta\vec{y})e^{i\phi(\Delta\vec{y})}$ is proportional to the Fourier transform of the normalized source intensity distribution\textemdash $I(\vec{x})=\langle |E(\vec{x})|^2\rangle/\iint \langle |E(\vec{x})|^2\rangle d^2x$, where $\vec{x}$ is the transverse position vector in the source plane\textemdash and depends on the displacement $\Delta\vec{y}=\vec{y}_2-\vec{y}_1$, between $\vec{y}_1$ and $\vec{y}_2$, and the difference of their distance from the optical axis, $r=|\vec{y}_2|^2-|\vec{y}_1|^2$:
\begin{equation}
    \label{eq1}
    v(\Delta\vec{y})e^{i\phi(\Delta\vec{y})}=e^{\frac{- i\pi r}{\lambda z}}\iint I(\vec{x})e^{\frac{-i 2\pi\vec{x}\cdot\Delta\vec{y}}{\lambda z}}d^2 x.
\end{equation}
Here $z$ is the distance from the source plane to the observation plane and $\lambda$ is the wavelength \cite{born2013,vanC,zernike1938concept,Goodman}. In interferometric imaging, measuring the complex visibility enables reconstruction of the source intensity distribution.

Our experimental setup demonstrating the GJC scheme is shown in Fig. \ref{fig2}. It consists of a pulsed heralded single-photon source and a pseudo-thermal light source in a single spectral-temporal mode mimicking filtered light from a star, which enables mode matching of the two sources. Both sources are derived from a titanium-sapphire (Ti:Sa) laser with \SI{80}{MHz} repetition rate, \SI{830}{\nm} central wavelength and approximately \SI{10}{\nm} full-width half-max (FWHM) bandwidth. To simulate starlight, i.e., a spatially incoherent light source, we direct a \SI{5}{\mW} laser beam onto a double-slit mask placed ~\SI{5}{\mm} in front of a rotating, nearly-Lambertian diffuser. The scattered light propagates \SI{1.02}{\m} to the detection plane resulting in a transverse coherence area on the order of a few square millimeters. A balanced, free-space BS separates the field into two spatial paths, enabling simultaneous coupling into two bare, polarization-maintaining, single-mode fibers (PM-SMF; Thorlabs PM780-HP) with a core diameter of \SI{5}{\um}, acting as small-aperture telescopes. One of the fibers, T1, is scanned in transverse position ($y_1$), to sample the field over a range of \SI{6}{\mm} in \SI{10}{\um} steps. The other fiber, T2, is statically positioned in the center of the scattered field. Scattered light from the rotating diffuser creates a time-varying speckle pattern in the far-field so that a single spatial mode exhibits thermal (Bose-Einstein) photon-number statistics (defined as $p(n)=\bar{n}^n/(\bar{n}+1)^{n+1}$) with average photon number $\bar{n}\approx0.008$ and second-order coherence $g^{(2)}(0)=2.00\pm 0.05$ \cite{estes1971}. Because the intensity distribution in the far field is spread over a region greater than \SI{10}{\cm}, each fiber collects the same mean photon number, $\bar{n}$, as long as the fibers are not displaced more than a few millimeters from one another. The joint state of the collected fields is described by a density matrix, $\rho_{ij}^{kl}$; with $i$ and $k$ labeling creation and annihilation operators for the field at T1, and $j$ and $l$ playing similar roles at T2. Further details throughout the paper are given in the Supplemental Materials \cite{supp}. 

The main beam from the Ti:Sa (\SI{1.24}{\W}) is frequency-doubled to a wavelength of \SI{415}{\nm} in a \SI{700}{\um}-thick beta barium borate (BBO) crystal to produce a \SI{100}{\mW} beam that pumps an \SI{8}{\mm}-thick potassium dihydrogen phosphate (KDP) crystal phase-matched for degenerate, collinear, type-II spontaneous parametric down-conversion (SPDC), which is designed to produce photons in the same spectral-temporal mode as the pseudo-thermal source. The source produces a two-mode squeezed-vacuum state that is nearly spectrally separable with a central wavelength of \SI{830}{\nm} where the herald mode has $\sim$\SI{5}{\nm} of bandwidth FWHM (horizontally\textendash H\textendash polarized), while the heralded mode has a bandwidth of $\sim$\SI{12}{\nm} FWHM (vertically\textendash V\textendash polarized) \cite{mosley2008}. A polarizing BS separates the H and V fields, where the V field is sent to a single-mode fiber connected to a superconducting nanowire detector (SNSPD). We measure an unheralded second-order coherence $g^{(2)}(0)=1.66\pm0.06$ (which implies a Schmidt number of $1.51$ and near separability \cite{Christ2011}) and an average photon number of approximately $0.01$, after correcting for detection efficiency. A detection event at the heralding detector indicates the presence of a single photon (or more) in the H-polarized beam. The density operator of the heralded field in the photon-number basis is approximately $\rho_{heralded}\approx\eta\, p(1)|1\rangle\langle 1|+(2-\eta)\eta\, p(2)|2\rangle\langle 2|$ where $\eta\approx 0.28$ is the measured heralding efficiency and $p(n)$ are thermal photon-number statistics, indicating the possibility of more than one photon in the pulse. 

The heralded H-polarized field is directed to a 50:50 BS realizing the PERS in Eq. \ref{eq}. A mirror bonded to a piezo-electric translator (PZT) in one path introduces a controllable, relative phase difference between the two paths. The PZT is controlled by a signal generator (SG) that applies a \SI{600}{Hz} triangular waveform, giving a time-dependent phase, $\delta(t)$, which varies between $0$ and approximately $4\pi$. The heralded PERS distributed to the telescopes is approximately given by Eq. \ref{eq}, which neglects the vacuum and higher-order photon numbers (a proof that the full state is entangled is given in the Supplemental Materials \cite{supp}). The two paths are coupled into PM-SMFs, indicated by S1 and S2 in Fig. \ref{fig2}.

The complex visibility of the simulated starlight, as defined in Eq. (\ref{eq1}), is measured by interfering the two modes of the PERS, S1 and S2, with the fields collected by the two fibers acting as telescopes, T1 and T2 in Fig. \ref{fig2}, at fiber BSs (FBS). The FBS outputs are sent to SNSPDs. All detection events are time-tagged using a time-to-digital converter. We keep coincidence events only when they are registered between different telescopes. The coincidence events,  occurring probabilistically at times $t_j$, can be associated with a known PZT phase, $\delta(t_j)$, by comparison to a phase-locked square voltage pulse used as a clock indicating the change in direction of the PZT. The set of event phases are used to estimate the complex visibility (see Supplemental Materials, \cite{supp} and \cite{tamimi2020} for details).

Theory predicts the estimated visibility measured by our experiment to be (see Supplemental Materials \cite{supp})
\begin{equation}
    \label{eq3}
    \tilde{v}(\Delta\vec{y})e^{i\tilde{\phi}(\Delta\vec{y})}\approx\pm \frac{2 p(1) \operatorname{Re}[\rho_{10}^{01}e^{i\delta}]\Gamma}{(2-\eta)p(2)\rho_{00}^{00}+p(1)(\rho_{10}^{10}+\rho_{01}^{01})},
\end{equation}

\noindent where $\pm$ is the phase difference acquired from the FBSs determined by which detectors register events, $\rho$ is the density matrix of the collected star photons that depends on Eq. \ref{eq1}, $p(n)$ is the Bose distribution of the two-mode squeezed state from the SPDC, $\eta$ is the heralding efficiency, $\Gamma\approx 0.5$ denotes the coherent overlap of the idler field mode from the SPDC and the field collected from the pseudo-thermal source. Given the experimental parameters, the predicted maximum value of $|\tilde{v}|$ from Eq. \ref{eq3} is $0.24$.
\begin{figure}
    \centering
    \includegraphics[width=\columnwidth]{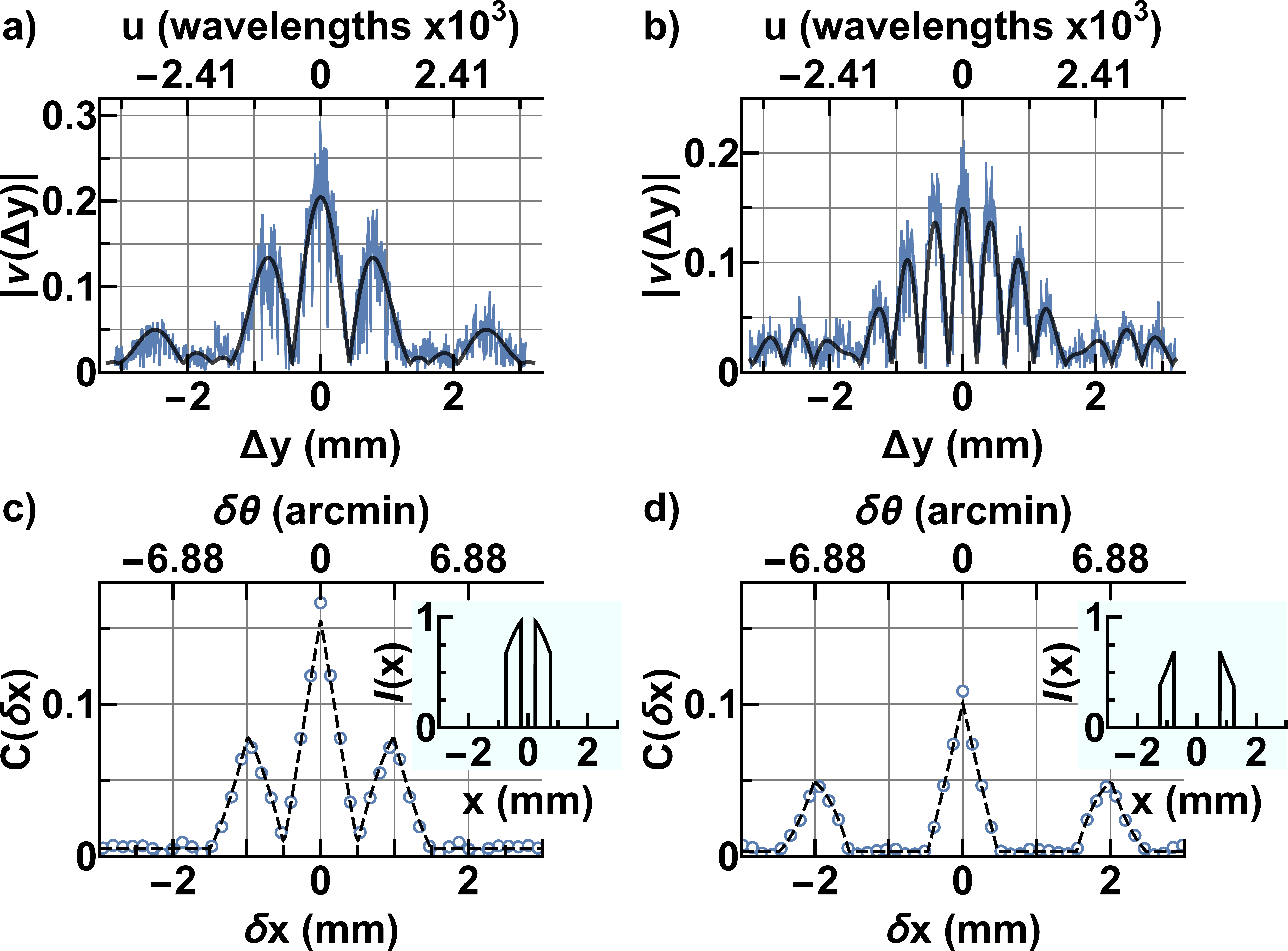}
    \caption{(a), (b) show the measured magnitude of the complex visibility, $|\tilde{v}(\Delta y)|$, versus telescope separation, $\Delta y$, in \SI{}{\mm} and, $u$, normalized by wavelength for double slits with \SI{1}{} and \SI{2}{\mm} separation, respectively. The points are linearly interpolated to guide the eye. The black line is a model fit that takes parameters found from the fits shown in (c), (d) and adjusts scaling and offset. (c), (d) show the reconstructed autocorrelation of the source distribution for the \SI{1}{} and \SI{2}{\mm} slit separations versus position in \SI{}{mm} and arcmin. Experimental error bars are subsumed by the circular points used (see Supplemental Materials \cite{supp} for details). The dashed black lines are the best model fits. Insets show the unnormalized source intensity distributions estimated from the model fit parameters.}
    \label{fig3}
\end{figure}

In this experiment, the magnitude of the complex visibility versus baseline is measured. The relative phase of the interferometer paths fluctuates on a \SI{10}{s} time scale, which is typically longer than necessary to acquire sufficient coincidence events to determine the amplitude and phase of the complex visibility at a given baseline, but does not allow a stationary phase reference between the measurements at different separations. However, the Fourier transform of the phase-independent, modulus-squared complex visibility as a function of baseline, $|v(\Delta\vec{y})|^2$, yields the autocorrelation of the normalized source intensity distribution, $C(\delta x)=\int I(x)I(x+\delta x)dx$ (see Supplemental Materials \cite{supp}). Such a reconstruction without interferometer stability is sufficient to verify a source distribution using a model fit.

For two known source distributions created from \SI{0.5}{\mm} wide vertically-oriented double slits separated by \SI{1}{\mm} or \SI{2}{\mm}, illuminated by a Gaussian-profile beam, Fig. \ref{fig3}(a) and (b) shows the magnitude of the complex visibility as a function of baseline in the horizontal direction, $\Delta \vec{y}$. Each data point in the measured visibility results from processing around 10,000 collected events\textemdash the PZT phases at the time of a coincidence, $\delta(t_j)$\textemdash over \SI{10}{s} (see Supplemental Details \cite{supp} and \cite{tamimi2020}). The Fourier transform of the squared visibility is proportional to the autocorrelation of the source intensity distribution, shown in Fig. \ref{fig3}(c) and (d). The estimated errors, which are equal in size to the circular points, account for statistical errors and phase fluctuations, but do not account for systematic errors. We also show a best fit of the reconstructed autocorrelation data to a model based on the autocorrelation of an intensity distribution describing the experiment (see Supplemental Materials \cite{supp}). Parameters in the model are the distance from the center of a slit to the midpoint between the slits, $d$; the width of the slits, $w$; and the radius of the common Gaussian beam illuminating the slits, $\sigma$.
\begin{table}[]
    \centering
    \begin{tabular}{c|c|c}
         & \SI{1}{\mm} Slits & \SI{2}{\mm} Slits\\
         \hline
         $\sigma (mm)$ & $1.678 \pm 0.271$ & $1.784 \pm 0.142$ \\
         \hline
         $d (mm)$ & $0.494 \pm 0.005$ & $1.006 \pm 0.004$\\
         \hline
         $w (mm)$ & $0.508 \pm 0.006$ & $0.476 \pm 0.006$
    \end{tabular}
    \caption{Autocorrelation fit parameters ($\sigma$, $d$, and $w$) for the double slits with \SI{0.5}{\mm} slit width, and slit separations of \SI{1}{\mm} ($d=\SI{0.5}{\mm}$) and \SI{2}{\mm} ($d=\SI{1}{\mm}$), which are illuminated by the same Gaussian beam.}
    \label{table1}
\end{table}

The best-fit parameters and errors determined using maximum likelihood estimation for the sources are shown in Table \ref{table1} and found to lie within fabrication tolerances of the slits ($\pm$\SI{20}{\um}). We also find that, as expected, the fitted Gaussian beam radius parameter in the two cases are equal within error. Overall, the fit to the expected intensity distributions for the two slits indicates that the GJC scheme is capable of source reconstruction. 

In conclusion, we have shown that an approximately single-photon state distributed across two paths can act as an entangled-state phase and amplitude reference for use in distributed VLBI, as proposed by GJC \cite{GJC}. The complex degree of coherence of a distant incoherent source was determined and served to reconstruct its intensity distribution. To achieve extended baselines, future work will study true thermal sources, reduce higher-order photon number contributions, and extend the scheme to detection of multiple spectral-temporal modes with the purpose of increasing detection rates. Lengthening baselines will require a quantum repeater chain to distribute the path-entangled light with lower loss than using optical fiber alone \cite{duan2001,sangouard2007}. Correcting for turbulence-induced phase distortion by the Earth’s atmosphere is possible in arrays with three or more telescopes using closure phase \cite{VLBIreview}, and progress toward such an implementation with PERSs can be found in \cite{diaz2021}. Further, it was proposed that a quantum network in which single-photon states could be stored in quantum memories and used on demand could provide a further increase in multiplexing ability \cite{khabiboulline2019I,khabiboulline2019II,Czupryniak,Czupryniak2022}. While functioning quantum networks are years in the future, if successful the proposed scheme could become one of the first practical uses of a quantum network and would open new horizons in visible-wavelength astronomical imaging. \nocite{ThermDiffState,SperlingVogel,PPT1,PPT2,saleh,walmsley1986experimental,ou2007,Variance,hurt}
\begin{acknowledgments}

This work was supported by the multi-university National Science Foundation Grant No. 193632 - QII-TAQS: Quantum-Enhanced Telescopy. We thank co-investigators Paul Kwiat, Virginia O. Lorenz, Yujie Zhang, Yunkai Wang, Eric Chitambar, and Andrew Jordan; as well as Tiemo Landes, Steven van Enk, Sofiane Merkouche, and Ramy Tannous for helpful discussions. 

\end{acknowledgments}
%

\clearpage


\section*{Supplementary material}

Supplementary Material

**Interferometric imaging using shared quantum entanglement**

Matthew R. Brown, Markus Allgaier, Valérian Thiel, John D. Monnier, Michael G. Raymer, and Brian J. Smith,

Here we provide several background and technical details that explain and support the methods used in the main body of the paper.

### A: Model of Coincidence Count Probability

To predict and model the coincidence count probability in the VLBI scheme, we need two important density matrices: one describing the 'faux' star electromagnetic field collected by the single-mode-fiber 'telescopes,' and the other describing the signal field from the down-converted pair, which is used as a path-entangled reference field (PRF). To model the density matrix describing the field collected by the telescopes, we first realize that the act of rotating the diffuser creates (by the central-limit theorem) Gaussian statistics in the complex field variables, as is well known in statistical optics [33]. Further, we know from the van Cittert-Zernike Theorem (Eq. 1 in the main text) that the fields collected by the telescopes are correlated [33]. We can then use the optical equivalence theorem (relating semi-classical field states to quantum states [30]) to write the density operator for the mutual state of light arriving at the telescopes as,

$$\hat{\rho}_{star} = \sum_{i,j,k,l=0}^{\infty} \rho_{ij}^{kl} (\hat{a}_1^\dagger)^i (\hat{a}_2^\dagger)^j |\{0\}\rangle\langle\{0\}| (\hat{a}_1)^k (\hat{a}_2)^l,$$

where $\hat{a}_m$ and $\hat{a}_m^\dagger$ are the annihilation and creation operators for mode $m$ and the density matrix, $\rho_{ij}^{kl}$, is given by

$$\rho_{ij}^{kl} = \begin{pmatrix} \rho_{00}^{00} & 0 & 0 \\ 0 & \rho_{10}^{10} & \rho_{10}^{01} \\ 0 & \rho_{01}^{10} & \rho_{01}^{01} \end{pmatrix}.$$

The density matrix elements are:

$$\rho_{00}^{00} = \frac{1}{1+\bar{n}_2+\bar{n}_1(1+\bar{n}_2-\bar{n}_2\,|v(\Delta y)|^2)},$$

$$\rho_{10}^{10} = \frac{\bar{n}_2(1+\bar{n}_1-\bar{n}_1\,|v(\Delta y)|^2)}{\left(1+\bar{n}_2+\bar{n}_1(1+\bar{n}_2-\bar{n}_2\,|v(\Delta y)|^2)\right)^2},$$

$$\rho_{10}^{01} = \frac{\sqrt{\bar{n}_1\bar{n}_2}\,v(\Delta y)\,e^{i\phi(\Delta y)}}{\left(1+\bar{n}_2+\bar{n}_1(1+\bar{n}_2-\bar{n}_2\,|v(\Delta y)|^2)\right)^2},$$

$$\rho_{01}^{10} = \frac{\sqrt{\bar{n}_1\bar{n}_2}\,v(\Delta y)\,e^{-i\phi(\Delta y)}}{\left(1+\bar{n}_1+\bar{n}_2(1+\bar{n}_1-\bar{n}_1\,|v(\Delta y)|^2)\right)^2},$$

$$\rho_{01}^{01} = \frac{\bar{n}_1(1+\bar{n}_2-\bar{n}_2|v(\Delta y)|^2)}{\left(1+\bar{n}_2+\bar{n}_1(1+\bar{n}_2-\bar{n}_2|v(\Delta y)|^2)\right)^2},$$

where, in the experiment, $\bar{n}_1 \approx 0.008$ and $\bar{n}_2 \approx 0.008$ are the average photon numbers collected by the telescopes T1 and T2 respectively, and $v(\Delta y)\,e^{i\phi(\Delta y)}$ is the complex interference visibility determined by the van Cittert-Zernike theorem (see Eq. (1) in the main text).

The intermediate state of the signal field post-selected on detecting a photon in the herald mode is approximated by

$$\hat{\rho}_{\text{Signal}} \approx \eta\, p(1)|1\rangle_H\langle 1|_H + (2-\eta)\eta\, p(2)|2\rangle_H\langle 2|_H,$$

where $\eta$ is the heralding efficiency (the probability that a photon in the herald mode heralds a photon in the signal arm) and $p(n)$ is the Bose distribution. The signal field is then propagated to a beam splitter and each output mode is then propagated one to each telescope to be interfered with the collected "faux" star field. As depicted in Fig. 1(b) in the main text, one mode is chosen to be have a phase change given by $\delta$. Using this state we can calculate the resulting probability to observe a coincidence event between pairs of detectors at different telescopes to be

$$P_{CC} = \eta_1\eta_2\eta\left((2-\eta)p(2)\rho_{00}^{00} + p(1)(\rho_{10}^{10} + \rho_{01}^{01}) \pm 2\,p(1)\,\text{Re}[\rho_{10}^{01}e^{i\delta}]\,\Gamma\right)$$

where $\eta_1, \eta_2, \eta$ are the detector efficiencies at telescope 1, telescope 2, and for the herald, and $\Gamma$ is the total mode overlap between the PRF and the thermal-like field collected by the telescopes. This form can be rewritten as

$$P_{CC} = \eta_1\eta_2\eta\left((2-\eta)p(2)\rho_{00}^{00} + p(1)(\rho_{10}^{10} + \rho_{01}^{01})\right)\left(1 \pm \frac{2\,p(1)\,\text{Re}[\rho_{10}^{01}e^{i\delta}]\,\Gamma}{(2-\eta)p(2)\rho_{00}^{00} + p(1)(\rho_{10}^{10} + \rho_{01}^{01})}\right),$$

which is similar to the classical expression for interference, where the first term is the total probability of a coincidence event, and the second indicates the coherent variation of probability of a coincidence event as either the visibility is changed by moving a telescope or the phase, $\delta$, is changed by varying one path length in the interferometer. The $\pm$ is determined by which output port of the beam splitter yielded the measured event. Thus, the measured complex visibility (and Eq. 3 in the main text) is given by

$$\tilde{v}(\Delta\vec{y})e^{i\tilde{\phi}(\Delta\vec{y})} \approx \pm\frac{2\,p(1)\,\text{Re}[\rho_{10}^{01}e^{i\delta}]\,\Gamma}{(2-\eta)p(2)\rho_{00}^{00} + p(1)(\rho_{10}^{10} + \rho_{01}^{01})}.$$

**B: Path Entanglement of the Heralded Field**

To demonstrate the entanglement of the heralded field, we must first notice that the general state (density operator) of the heralded field including loss and non-unit efficiency of the heralding detector is given by thermal difference states [42], represented by the form:

$$c\sum_{n=0}^{\infty}(q^n - d(pq)^n)|n\rangle\langle n|$$

where $c, q, d$, and $p$ are known functions of the average photon number of the initial down conversion, the quantum efficiency of the heralding detector, and the loss the heralded field experiences prior to the beam splitter (which we assume to be upper bounded by 10%) and are defined in [42]. To verify entanglement between the fields in the two paths after the heralded field is split by a beam splitter, we show that the projection to the qubit subspace of the two Hilbert spaces defining the bipartite state is entangled, which is sufficient to show the entire state is entangled [43]. To show the entanglement of the qubit subspace, it is sufficient to show that the partial transpose of the density matrix is not positive by, for instance, determining that the determinant is negative (implying that at least one of the eigenvalues is negative) [44,45]. The partial transpose of the qubit density matrix, derived from the state above, is

$$c\begin{pmatrix} 1-d & 0 & 0 & \frac{-i(q-dpq)}{2} \\ 0 & \frac{q-dpq}{2} & 0 & 0 \\ 0 & 0 & \frac{q-dpq}{2} & 0 \\ \frac{i(q-dpq)}{2} & 0 & 0 & \frac{q^2-d(pq)^2}{2} \end{pmatrix}.$$

Hence, the determinant is

$$\frac{c^2(q-dpq)^2}{4}\left(\frac{c^2(1-d)(q^2-d(pq)^2)}{2}-\frac{c^2(q-dpq)^2}{4}\right).$$

For our measured experimental parameters and an estimated upper bound on the loss (10%) from the down conversion crystal to the beam splitter ($c \approx 351.38$, $d \approx 1$, $p \approx 0.72$, $q \approx 0.009$), we find the determinant to be $\approx -0.038$ thus showing that the state is entangled.

### C: Aligning Telescope Fibers and Measuring Scattered Light

One important aspect of the experiment is alignment of the telescope fibers so that they collect, approximately, the same transverse part of the field. To first order, we first try to find the peak of the intensity distribution (usually Gaussian) without the diffuser since each fiber is much smaller than the spatial distribution in the far field. However, this method becomes less precise as the intensity distribution at the diffuser becomes larger. The only way to truly center the fibers, then, is to raster scan and search for interference fringes in the coincidences as described by the van Cittert-Zernike theorem.

Further, we also need to ensure that the light scattered by our diffuser has properties consistent with a pseudo-thermal source and has no ballistic throughput. With our fibers aligned, we measure a $g^{(2)}(0) = 2.00 \pm 0.05$, consistent with thermal-field statistics. However, a $g^{(2)}(0)$ value alone does not ensure that we truly have a thermal state; therefore, we image, on a camera (Thorlabs DC1545M), the spatial structure 48.26 cm away from the diffuser and near the center of the transverse structure shown in Figure B.1a). From this image, we gain two major things: first, we have good spatial scattering without a ballistic component; second, we can use the image to make a histogram of the instantaneous intensity values, as shown in Figure B.1b). For near-single-mode thermal-like light, we expect this form to be a negative exponential for large intensities with a peak at low intensities due to the nonzero size of the individual pixels [46,47]. Finally, we want the source coherence area to be as small as possible. To test for this, we first realize that there is a generalized van Cittert-Zernike theorem that deals with a nonzero coherence area at the source [33]. If the coherence area were too large, we would notice the intensity distribution in the far field to have a small angular spread, that is, to not be Lambertian. We measured that the intensity angular spread indicates a source coherence roughly less than 6 microns, small enough to be considered point-like when examining the far-field correlations only near the beam axis.

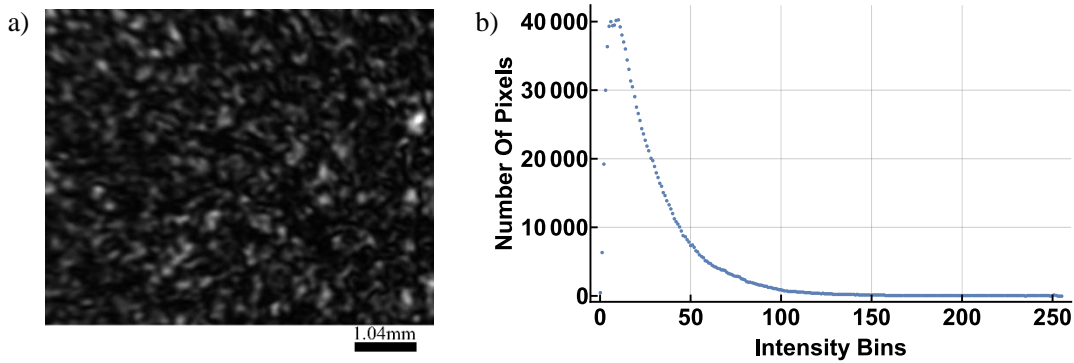

Figure B.1. Scattering of the light through the diffuser. In a) we show a far-field camera image of the scattered light. Note the lack of a ballistic component (bright center spot) in the scattering. In b), we show the histogram of the number of pixels sharing a specific intensity captured by the camera. This shows the nearly Gaussian and Lambertian character of the scattering medium.

### D: Timing and Synchronization

In order for the non-local oscillator field and the collected thermal-like field to interfere at each of the fiber beam splitters, the temporal envelopes of the pulses must overlap. To accomplish maximum overlap, Hong-Ou-Mandel (HOM) interference is observed between the unheralded single photon and the effective thermal

field collected by the telescopes. There are two delay stages to control the relative timing between the 100 fs, thermal field pulses and the unheralded SPDC 100 fs pulses. The stages are then set to the position of maximum two-photon interference [48]. Once the delays are set to the position maximizing the HOM interference, the experiment is ready to record data. Below are some technical details covering the synchronization processing of the time tags collected during measurement.

The data comes in the form of time-tags with a timing jitter of 13 ps from two ID900 time-to-digital converters (TDC). The "start measuring" command from the controlling computer had a different time of arrival between the two TDCs, hence, they must be synchronized for coincidences to be detected. Specifically, this task is to find the timing offset between the two TDCs and add this time to the time tags from the TDC that received the start message first. To measure the absolute timing delay between the TDCs we use a 600 Hz square-pulse voltage source, phase-locked to the 600 Hz triangle-wave signal that drives the PZT, which makes a time-varying phase between the different paths of the single photon. To make the square pulse common between the two TDCs, it is routed through a delay generator (SRS DG535) where the signal is split into two channels, one sent to each TDC. Any differing delays accrued by the cable lengths were accounted for by measuring the timing difference on a single TDC. In our case, the timing delay was much less than the 6 ns coincidence-window (<100 ps) and is negligible. Given that each TDC has a common signal, we note that during the internal phase-locking procedure of the square-pulse generator there was a slight timing change between subsequent square pulses as the square pulse is phase-locked to the triangle wave. Since each TDC will see this sudden shift in timing, we used this shift to calculate the absolute timing between the TDCs by the following procedure: From one TDC, using the stream of time tags resulting from the square wave, calculate the inverse of the difference of subsequent time tags (hereafter called the frequency). The list of frequency values can be searched for any frequencies that are different (by usually more than a few Hz; a threshold can be applied by measuring the standard deviation of these arrival frequencies of the square wave and is in the sub-Hz regime). The last position where the frequency deviates from 600 Hz was selected. Since the frequencies are made of two consecutive time tags ordered by their time-of-arrival (first and last), we took the time tag directly after the last time tag of the pair to ensure that there are no further drifts in frequency due to the phase-locking procedure. The resulting time tag (called the synchronization time) at each TDC are compared and the difference of these time tags yields the absolute timing difference between the two TDCs to the nanosecond level or better. To confirm synchronization, coincidences between the herald channel on one of the TDCs and one of the detection channels (which contains a single photon from the other half of the PDC pair) connected to the other TDC are measured. Finally, all time tags prior to the synchronization time were dropped since the square pulse was not yet phase locked to the triangle wave driving the PZT.

**E: Determining Complex Visibility from Time Tags**
The time tags produced by measurement are processed in software. In our case, the information is contained in the probability of a coincidence as found in section A. We must then take each of the measured time-tags and compute the coincidences, with a 6 ns window, between the detectors of interest including the herald detector. There are four sets of time-tags corresponding to the four possible detector combinations that give information (the other two possible groupings, coincidences at the same telescope, cause the loss of a factor of two in data rate, as explained in [10]). Furthermore, two of the four possible combinations exhibit a π-phase shift compared to the other two. To make sure that the phasors from all four arrangements are in phase, an extra π-phase is added to the phases measured from one of the arbitrarily chosen, but constant, pairs of detector combinations. The heralded data is then created by finding coincidences (again with a 6 ns window) with the herald signal from PDC and each of the four coincidence streams found previously. It is important to note that all data currently is still in time-stamp format.

Each coincidence event is associated with a specific PZT phase found by reference to the known triangle waveform. This phase ($\delta_j$), also called a phase tag, can be determined by the known function applied by the signal generator that drives the PZT (600 Hz triangle pulse with ~4π amplitude) given by the time difference of the coincidence time-tag to the time-tag of the nearest square pulse, which is phase-locked so that its rising edge starts at the beginning of a triangle wave (see section B). Effectively, these square pulses also act as a clock. The amplitude of the triangle pulse is measured by aligning a classical laser beam into the same paths

(and fibers) as the distributed single photon. By then interfering the outputs of the single-photon collection fibers using a fiber beam splitter and measuring the interference pattern of the output as the PZT is swept over time (by e.g. a photodiode and analog-to-digital converter), one can estimate the approximate phase applied by the PZT through curve fitting. The peak-to-peak voltage applied to the PZT is adjusted so that a relative phase between the two paths of roughly $4\pi$ is applied to the optical fields.

We estimate the measured complex visibility in a single trial recording many events by averaging over the resulting phasors to yield the mean phasor,

$$2 \langle e^{i\delta_j} \rangle = \tilde{v}(\Delta y) e^{i\tilde{\phi}(\Delta y)}. \tag{12}$$

The brackets indicate averaging over $N$ samples, where $N$ is of the order ten thousand. This technique is called phase-tagged photon counting [38]. This estimator for the complex visibility can be shown to be unbiased.

### F: Phase Fluctuations, Autocorrelations, and Error Propagation

Environmental phase fluctuations in our system occur on an estimated times scale of 10 s. Given that the phase fluctuations may be of order of $2\pi$, we cannot currently achieve full imaging as defined in the van Cittert-Zernike theorem. Rather, we use the fact that the autocorrelation, $C(\delta \vec{x})$, of the source intensity is related to the squared modulus of the measured visibility [33] through

$$C(\delta \vec{x}) = \iint I(\vec{x}) I(\vec{x} + \delta \vec{x}) d^2 x = \iint |v(\Delta \vec{y})|^2 e^{-i \frac{2\pi \delta \vec{x} \cdot \Delta \vec{y}}{\lambda z}} d^2 \Delta y.$$

Thus, we are interested in estimating quantities related to the magnitude of the visibility.

In theory, each trial should yield a complex value for the mean phasor that is normally distributed with equal variance in the real and imaginary axes, with no correlation between them, and centered around the complex visibility by the central limit theorem. However, due to the uncontrolled environmental phase fluctuations between trials, the complex distribution may be more complicated. Thus, a method of determining the variance in the complex direction of the mean value (our estimate for the visibility) is needed. This is achieved by rotating the measured distribution (of approximately 10,000 complex samples) such that the new mean phasor points along the positive real axis. One can then compute the variance in the real direction, which corresponds to the variance of the original distribution in the complex direction of the visibility [49]. With this definition, one can use standard error propagation from $|v(\Delta \vec{y})|^2$ to $C(\delta \vec{x})$; however, propagation from $v(\Delta \vec{y})$ (in the direction of $v(\Delta \vec{y})$) to $|v(\Delta \vec{y})|^2$ requires the use of a second-order delta method for cases where the mean value is small [50]. Given that our error-propagation function, $f(x) = x^2$, is a polynomial of degree two, this error propagation is exact. In particular, this means that

$$n(f(X_n) - f(\mu)) \xrightarrow{D} \sqrt{n} \, N(0, f'(\mu)^2 \sigma^2) + \frac{\sigma^2 f''(\mu)}{2} \chi_1^2.$$

Here, $n$ is the number of samples, $N(\mu, \sigma^2)$ is the normal distribution centered at $\mu$ with variance $\sigma^2$, $\chi_1^2$ is the chi-squared distribution with one degree of freedom, and $\xrightarrow{D}$ symbolizes convergence in distribution. So, the new mean and variance are given by $\mu^2 + \sigma^2$ and $4\mu^2 \sigma^2 + 2\sigma^2$ respectively, where $\mu$ is the mean of $v(\Delta \vec{y})$ and $\sigma^2$ is the asymptotic variance (i.e. variance due to the central limit theorem) of $v(\Delta \vec{y})$ in the direction of the mean. Both $\mu$ and $\sigma^2$ can be determined directly from the sampled distributions. This procedure yields error estimates that are smaller than the circles representing data points in the reconstructed autocorrelation functions in the main text, Fig. 3.

### G: Model Fitting

In this paper, we fit the inferred source intensity autocorrelation function $C(\delta x)$ to a theoretical model that represents well the known source distribution. Here we detail the model and the fitting parameters we assume,

along with the fitting parameters for the 1 mm and 2 mm slits. We define the model for the normalized intensity distribution as

$$I(x;\sigma,w,d) = \frac{\sqrt{2/\pi}}{\sigma\left(\text{Erf}\left(\frac{2d+w}{\sqrt{2}\sigma}\right) - \text{Erf}\left(\frac{2d-w}{\sqrt{2}\sigma}\right)\right)} e^{\frac{-2x^2}{\sigma^2}} \left(\text{UnitBox}\left(\frac{x-d}{w}\right) + \text{UnitBox}\left(\frac{x+d}{w}\right)\right),$$

where $\sigma$ is the radius of the Gaussian beam that impinges on the double slit, $w$ is the width of an individual slit, $2d$ is the distance between the centers of the two slits, and the $\text{UnitBox}(x)$ function is defined as 1 when $|x| \leq \frac{1}{2}$ and 0 otherwise. We can then calculate the autocorrelation of $I(x;\sigma,w,d)$, which is omitted here due to its complexity. However, there are a few complications that must be accounted for: the visibility is not unity in our experiment, the discrete nature of the data, and the non-zero nature of the resulting data. The first two of these issues require us to use a multiplicative scaling parameter, $A$, while the second forces an offset parameter, $B$. Given that we need a multiplicative scaling factor, we do not need to worry about normalization. Hence, we will use the unnormalized intensity

$$\tilde{I}(x;\sigma,w,d) = e^{\frac{-2x^2}{\sigma^2}} \left(\text{UnitBox}\left(\frac{x-d}{w}\right) + \text{UnitBox}\left(\frac{x+d}{w}\right)\right).$$

Thus, $C(\delta x) = \iint \tilde{I}(x;\sigma,w,d)\tilde{I}(x+\delta x;\sigma,w,d)d^2x$ and the final model is given as

$$A\, C(\delta x) + B,$$

where we fit to all parameters listed above using Mathematica's NonlinearModelFit function. The full parameter list of the fits found for the data is shown below.

|   | 1 mm Separated Slits | 2 mm Separated Slits |
|---|---|---|
| $\sigma$ | $1.678 \pm 0.271$ mm | $1.784 \pm 0.142$ mm |
| $d$ | $0.494 \pm 0.005$ mm | $1.006 \pm 0.004$ mm |
| $w$ | $0.508 \pm 0.006$ mm | $0.476 \pm 0.006$ mm |
| $A$ | $211.18 \pm 24.69$ | $353.79 \pm 68.99$ |
| $B$ | $0.00522 \pm 0.0001$ | $0.003 \pm 0.00006$ |

The beam radii inferred are comparable; and, the larger error is attributed to the slight differences in the curvature of the side peaks and the necessity to fit a scaling parameter.

To fit the measured visibility, we use these parameters in the Fourier transform of $\tilde{I}(x;\sigma,w,d)$ and allow for scaling and offset parameters as usual. Below, we show the functional form of the modeled absolute value of the visibility since the phase is unimportant in this experiment:

$$|v(\Delta y; A', B', \sigma, w, d)| = \left|\frac{A'}{2} e^{-\frac{\pi^2 y^2 \sigma^2}{2\lambda^2}} \sqrt{\frac{\pi}{2}} \sigma \left(\text{Erf}\left[\frac{2d\lambda + w\lambda - i\pi y \sigma^2}{\sqrt{2}\lambda\sigma}\right] - \text{Erf}\left[\frac{2d\lambda - w\lambda + i\pi y \sigma^2}{\sqrt{2}\lambda\sigma}\right] + \text{Erf}\left[\frac{-2d\lambda + w\lambda + i\pi y \sigma^2}{\sqrt{2}\lambda\sigma}\right] + \text{Erf}\left[\frac{2d\lambda + w\lambda + i\pi y \sigma^2}{\sqrt{2}\lambda\sigma}\right]\right)\right| + B'.$$

These expressions allow us to propagate the model fit of the autocorrelation back to our measured visibility.


\end{document}